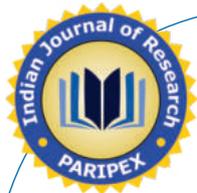

# ORIGINAL RESEARCH PAPER

## A SURVEY: INFORMATION SEARCH TIME OPTIMIZATION BASED ON RAG(RETRIEVAL AUGMENTATION GENERATION) CHATBOT.

**Machine Learning**

**Jinesh Patel**

**Arpit Malhotra**

**Ajay Pande**

**Prateek Caire**

**ABSTRACT**

Retrieval-Augmented Generation (RAG) based chatbots are not only useful for information retrieval through question-answering but also for making complex decisions based on injected private data. we present a survey on how much search time can be saved when retrieving complex information within an organization called "X Systems" (a stealth mode company) by using a RAG-based chatbot compared to traditional search methods. We compare the information retrieval time using standard search techniques versus the RAG-based chatbot for the same queries. Our results conclude that RAG-based chatbots not only save time in information retrieval but also optimize the search process effectively. This survey was conducted with a sample of 105 employees across departments, average time spending on information retrieval per query was taken as metric. Comparison shows us, there are average 80-95% improvement on search when use RAG based chatbot than using standard search.

## INTRODUCTION

Before conducting this survey, previous studies indicated productivity improvements in the healthcare sector through the use of chatbots [1]. However, the method faced challenges due to the complexity of data collection. Modern chatbots are no longer limited to text-based interactions; they have evolved into multimodal systems capable of processing text, images, and videos [2]. With the rapid advancements in generative AI (GenAI), chatbots not only enhance productivity but also serve as innovative tools for information mining. In this survey, we measured task completion time by comparing a traditional information retrieval system, such as a search engine, with a chatbot in real-world scenarios, such as within a corporate environment. Before delving into the detailed survey findings, we explain how we implemented the chatbot using key components. Following this, we outline the survey methodology and discuss the impact of chatbot-assisted information retrieval on productivity, which was the primary goal of this survey. The survey was conducted at a stealth-mode startup focused on improving agricultural production through AI and IoT technologies. Due to its stealth status, the company's name and specific products cannot be disclosed. The survey encompassed various departments, including software engineering. The primary source of information sharing within the company was the Confluence system, making it the focus of our data analysis. The survey examined queries of varying complexity and length across departments. For instance, a straightforward query for the sales department might be, "What was last quarter's revenue?" whereas a more complex query could be, "What if we increase the sales price by x% for customer group A? How much revenue impact can we predict for next year?" These complex queries often required searching through multiple Confluence pages and performing manual calculations. Using the chatbot, such queries could be answered within seconds, significantly reducing the effort required. This reduction in time and effort was measured as a productivity improvement. Although implementing a Retrieval-Augmented Generation (RAG) system was not the primary focus of this paper, we leveraged OpenAI's API for generating responses and Milvus as the database for vector data storage. We processed approximately 2,300 pages of Confluence data. While there is room for improvement in the chatbot implementation, our focus remained on evaluating productivity gains. A high-level implementation architecture is shown in the figure below in survey and methodology section.

## Survey Design and Methodology

Below figure explained components how we implemented RAG chatbot for information finding on platform like confluence. If you are not aware with confluence, it is the popular document management system developed by company called "Atlassian". It gained popularity because of easy document management and project management in IT industries. Again, for this survey our focus is not to build a robust chatbot but rather its application in general answers in document search. In current era, chatbots are built from pretrained LLM like LLMA2 [3] from step 1. Since main objective of this paper is not to implement RAG chatbot, we may enhance this implementation by selecting optimal algorithm for embeddings etc. but this was out of the scope of the paper. Now, let's move to main objective of the paper, How we designed survey and survey methodology.

**Participants and Survey Instrument**

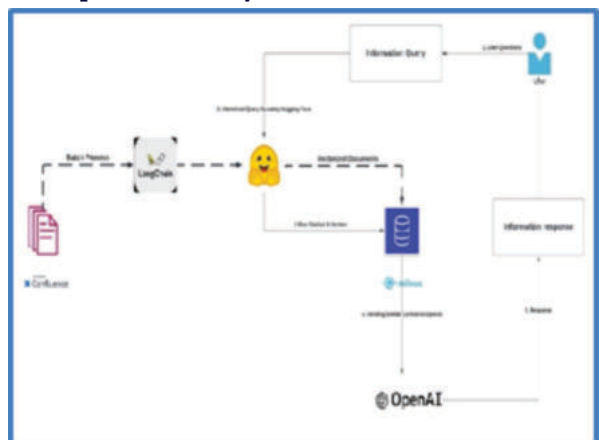

**Figure 1. Building Blocks of RAG Based Information Chat Bot.**

In this project, we have built RAG based chatbot by using basic predefined building blocks. These chatbots are deployed to mission critical systems and estimated LLM market sized is 4.35 billion US$ [4]. We selected this design for several reasons. 1) Most of the components are open source. although OpenAI charges per API call, it is relatively affordable and sustainable. 2) Simple Integration, All components showed in figure are integrated very well. Now let me explain how this system works in the brief before explaining survey design and methodology. You can see there are two lines connected to components. Dotted line

represents passive or batch process. Solid line represents active or user process. First, From the targeted confluence documents ( from within we designed our survey questions or wants to retrieved information), we connected those document via LangChain plugin and convert into vector by using simple ADA 002 vectorization. The main advantage of this vectorization is, it has 1536 dimensions that covers most of the semantics of our total document size of 2300 pages and well works with OpenAI's eco-system as it was developed by OpenAI. Once Documents converts into vector, we insert these vectors to well-known vector database MilvusDB.

MilvusDB is highly scalable, robust and versatile database just performing vectors operations like find various similarities like cosine, Manhattan or Euclid. Once, we have vectorized data into db system, we will be using the same embeddings ( ADA 002) to convert our query. We provided Web UI to send query to our system. User can generate query to retrieve information via this interface. The system works following steps.
1. User generates query for information that interested or finding answer from predefined confluence pages.
2. Query is vectorized by using the same embeddings (ADA 002) with the help of huggingface interface.
3. Vectorized query sent to milvus db for finding similar vectors (that stored already from confluence pages).
4. Collected similar vectors are treats as context of query and send to OpenAI's API as prompt.
5. Once, we got answer back from OpenAI, we treats as answer and display to user.
6. Follow up questions or queries, we append to context and repeat

Main object of this paper is to show, how RAG based chatbots improves information retrieval from complex document systems like confluence. For this we designed survey to stealth mode startup covering various departments and various range of people with essential to expert level of IT knowledge. Survey covers majorly five departments with 105 employees. These five departments are engineering ( covers software engineering, data science, QA and product management), marketing, sales, HR. Person from age twenty to fifty five years participated in the survey. Job wise, intern to company's executive were targeted participants. Demographic information was irrelevant for this survey as it needed only sufficient professional knowledge of English language. Male to Female ratio were kept 1.5:1 as there was skewed in survey population. For this survey, we asked randomly selected departmental related queries to search on confluence system and the same queries to be asked in RAG based chatbot. Of course, truth of the source kept the same and that we asked from confluence system. chatbot data was injected from this same documents Queries were vary from simple to complex. For example, "what is total sales in 2023?" or more complex and specific like "what is campaign cost for customer A and return investment on the same customer in April 2024?". For engineering department, query like "where is design code for Kubernetes infra?" or "how to debug specific API service on google cloud?" of course, all these queries answers were available in confluence document. We have divided queries based on length and complexity. For simplicity, we did not perform semantic analysis on query to determined complexity. However, we assume larger queries have more complexity and based on this we performed the survey.

### Survey Procedure
After we decided who to be involved in the survey and what needed to be asked. We decided on survey measurement. We measure time via Focus timer app on MAC window desktop. We gave mix of 10 queries length of queries vary to 5 - 50 words and asked people to find answers from confluence and measure time. The same set of queries were asked to find answer on chatbot and record the time. This survey was done in person and the basic unit of measurement was time in seconds to find answer or information in both the methods ( via regular search and chatbot). During the survey, we asked participant to stay focus while finding information. some of the queries might needed manual calculation like average or frequency count. None of the queries were asked from outside confluence document scope.

The Average words for query was 10.5 ( major queries were finding single informative answer like "what is revenue generated from customer A in previous year?". As we mentioned earlier, some of the queries need manual calculation, for example "what is percentage sale increment for customer B during previous two years?".The same queries were also, handled by chatbot and calculation were done automatically by chatbot.

### Analysis Method
Information retrieval time based on asked queries was a main performance measurement. We measured time to get information with personal effort like search on document and manual calculation and measured time to get answer for the same information with chatbot. Of course, we measured the truthfulness of answer as well when we retrieve information from chatbot since chatbot might hallucinate [5].

After measuring time for information retrieval, we performed statistical analysis to determine performance improvement. This analysis involved time to take complete all ten queries, average time to complete single query. Average time taken to complete longest ( complex) query and average time taken to complete the shortest query. We measured this data point across the department, with age bin and then we made conclusion out of these data points.

### Related Work
RAG and GenAI have been hot topics nowadays, and many researchers are targeting this space in surveys to compare and analyze developments, while some research works emphasize the standardization and certification process of GenAI-based chatbots[6]. In this survey[7], the authors conducted a survey on university students about their awareness of general GenAI-based chatbots and found that most of the students found them beneficial. Some surveys have been conducted to find the overall harms and benefits of AI[8] this international survey provides insight into how other countries approach AI and the regulation of related technologies by providing an overview of the national AI strategies of 26 countries. Apart from GenAI-based chatbots, most surveys conducted from a usability and ethics perspective in the AI space. For example, this survey[9] was conducted on the ethical principles of AI and its implementations, concluding that ethical principles need to be integrated at every stage of the AI lifecycle to ensure that the AI system is designed, implemented, and deployed in an ethical manner.

Some surveys are based on GenAI usability in niche fields like medicine and telecommunications. In one survey on communication about 5G, it was concluded that a fully operative and efficient 5G network cannot be complete without the inclusion of artificial intelligence (AI) routines[10]. Also in the same communication field, this survey explains the challenges of AI-enabled 5G networks[11]. Similarly, the New Generation of AI (NGAI) and its applications in power system operation have become research hotspots[12].

GenAI is not only used in technical and medical fields but also in fields like fashion. This survey explains how fashion applications are benefiting greatly from the development of machine learning, computer vision, and artificial intelligence[13]. It also mentions that problem formulations, method comparisons, and evaluation metrics are illustrated for each topic of fashion research.

Finally, some researches closely related to GenAI technology itself. This survey[14] questions ChatGPT's security, privacy, ethical, and legal challenges. One survey found is the closest related to this paper[15], explaining about a variety of chatbots for different services. However, most of the focus of this survey was on the design techniques for building the chatbot depending on the services meant to provide for the users. In this paper, we are more focused on users' time optimization. None of the above studies were conducted to measure a constructive time optimization approach on daily productivity increment and time optimization for tasks like searching content in complex data systems like Confluence.

**RESULTS AND DISCUSSION**
In this section, we are discussing our survey result and analyze how this survey optimized search time. As we mentioned earlier, we built chatbot by using approx. 2300 pages data on confluence and conducted survey of five departments, we first see how each department was spending search time to get answer of simple queries, most of the queries did not involve any manual calculation and they had straight forward answer in confluence. for example, One of the example of the queries asked in sales department was "what is total sale of our company in 2023?", of course, the page was dedicated to sales in sales department page. However we saw significant time improvement while using chatbot. First below table mentions about how survey responder distributed by department.

**Table 1: Total Participants by Departments.**

| Sr. | Department name | Total survey responder | Notes |
|---|---|---|---|
| 1 | Sales | 25 | |
| 2 | Marketing | 20 | |
| 3 | Human Resource | 5 | |
| 4 | Engineering | 45 | |
| 5 | Executives | 10 | Cxx People |
| Total | | 105 | |

For the simple queries we defined as they are straight forward and easy to find on confluence, we found below result based on departments. We asked 10 simple questions and record time to measure on finding on confluence and the same set of questions we asked people to find via chatbot. We took time to find answers and took average in seconds per person. Below table are our findings.

**Table 2: Average Time Spent by Department to Find Simple Queries.**

| Sr | Department name | Average time to find 10 queries on confluence (in sec) | Average time to find 10 queries in chatbot (in sec) |
|---|---|---|---|
| 1 | Sales | 690 | 130 |
| 2 | Marketing | 750 | 140 |
| 3 | Human Resource | 650 | 120 |
| 4 | Engineering | 550 | 110 |
| 5 | Executives | 620 | 150 |
| Total | | 3260 | 650 |

Above table shows, we have reduced query searching time via chatbot around 80% of the time. This is because responders are familiar with the query topic and no complex calculation were involved. We also conducted another survey with the complex queries( avg length 15 words), where it needed multiple page scanning on confluence. We found result below, optimization by chatbot increases as well. We also conducted survey on age bin with simple queries. Results are displayed below.

**Table 3: Average Time Spent by Department to Find Complex Queries.**

| Sr | Department name | Average time to find 10 queries on confluence (in sec) | Average time to find 10 queries in chatbot (in sec) |
|---|---|---|---|
| 1 | Sales | 2280 | 245 |
| 2 | Marketing | 2590 | 315 |
| 3 | Human Resource | 2150 | 280 |
| 4 | Engineering | 1850 | 170 |
| 5 | Executives | 2400 | 180 |
| Total | | 11270 | 1190 |

The total average time shows around 90% of improvement on query search involved multiple confluence page scanning. last survey we conducted that not only involved multiple confluence pages to find answer but also involves manual calculations. The result displays in below tables

**Table 4: Average Time Spent by Department to Find Complex Queries that Involves Manual Calculations.**

| Sr | Department name | Average time to find 10 queries on confluence (in sec) | Average time to find 10 queries in chatbot (in sec) |
|---|---|---|---|
| 1 | Sales | 3850 | 250 |
| 2 | Marketing | 4160 | 350 |
| 3 | Human Resource | 4050 | 300 |
| 4 | Engineering | 3500 | 200 |
| 5 | Executives | 3800 | 250 |
| Total | | 19360 | 1350 |

The average time shows around 95.5% improvement while using chatbot to find very complex queries that involves manual calculations. Since chatbot can automatically do such a calculations, the time optimization is the highest in this scenario.

Future more, As we increased the complexity and hardness of the query, chatbot solves effectively with more time optimization.

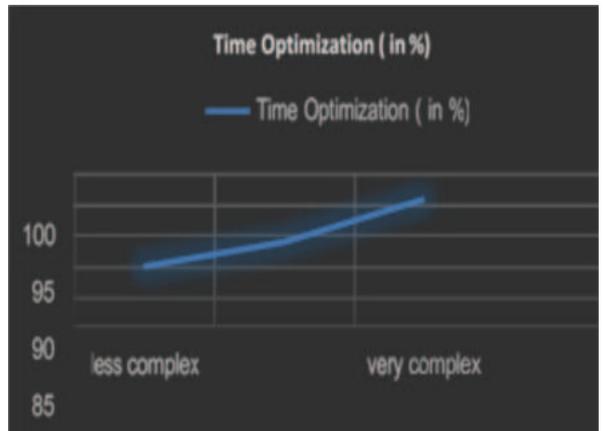

Figure 2. Query Time Optimization vs Query Complexity.

Based on our conducted survey, we can correlate query complexity and search time optimization while using chatbot. When query complexity increased ( i.e query length, query semantics), the time

**Table 5: Average Time Spent by Age Bin to Find Simple Queries.**

| Sr | Age bin ( in years) | Average time to find 10 queries on confluence (in sec) | Average time to find 10 queries in chatbot (in sec) |
|---|---|---|---|
| 1 | 20-30 | 590 | 105 |
| 2 | 30-45 | 650 | 120 |
| 3 | 45-55 | 680 | 150 |
| 4 | 55-60 | 900 | 180 |
| 5 | 60 and above | 1020 | 210 |
| Total | | 3840 | 765 |

We also plot based on the age group, how much manual search query time increases vs chatbot time increase.

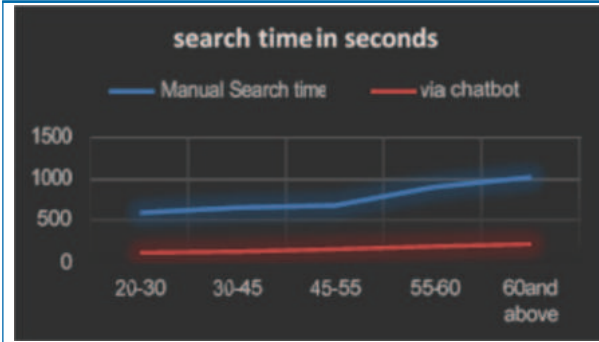

**Figure 3. Query Time by Age Bin (Chatbot vs Manual Search)**

This shows linear increment in search time while using chatbot by age group however manual search is not a linear.

## CONCLUSION

This survey was conducted on a stealth mode startup to study the effect of information retrieval via a RAG-based chatbot. The study shows that the chatbot has a positive contribution to documentation search in systems like Confluence. Overall, the usage of the chatbot saves 80- 95% of man-hours across the organization, regardless of a person's ability to use effective search techniques. It also concluded that as query complexity increases, manual information retrieval time increases polynomial, while chatbot information retrieval time remains almost linear. Additionally, search time is directly proportional to age when using manual search, unlike with the chatbot, where the time is almost constant.

The use of chatbots in information search and question answering is obviously beneficial to organizations. However, bundling chatbots with data privacy remains a concern for any organization. Since chatbots use external APIs (in this case, OpenAI), they are sending data to a third- party organization. Organizations should be aware of this fact and should not build chatbots on data where privacy is involved, like HIPAA or GDPR compliance. Finally, chatbots and other GenAI technologies are prone to inaccuracies and false results. Even sophisticated systems like ChatGPT provide warnings about this. Therefore, whatever result or answer we obtained via the chatbot in this project, we verified before accounting for it. In this survey, the probability of hallucinating was negligible as the data was not enormous. However, those who build such systems on huge datasets should be aware of this fact.

### Future Research
We conducted a survey on a stealth mode organization to assess the efficiency of information retrieval. The audience for this survey was limited, and it did not cover many samples. We could conduct such a survey on a larger organization to see the impact and bolster the results. Furthermore, there are more fields like image analysis and medical report analysis where we can compare manual efforts versus chatbot analysis to see the impact on time optimization